\documentclass[aps,prd,twocolumn,superscriptaddress,showpacs,letter,nofootinbib]{revtex4-1} 
\usepackage[utf8]{inputenc}
\usepackage{graphicx}
\usepackage{dcolumn}
\usepackage{bm}
\usepackage{cancel}
\usepackage{color}             
\usepackage{epsfig}
\usepackage{amssymb}
\usepackage{amsmath}
\usepackage{multirow}
\usepackage{hyperref}
\usepackage{cleveref}
\usepackage{lineno,color}
\usepackage{epstopdf}

\begin{document}

\title{Semihard interactions at high energies}
\thanks{Presented at ``Diffraction and Low-$x$ 2024'', Trabia (Palermo, Italy), September 8-14, 2024.}

\author{T.~V.~Iser}
\email{thomas.iser@ufrgs.br}
\affiliation{Instituto de F\'isica, Universidade Federal do Rio Grande do Sul, Caixa Postal 15051, CEP 91501-970, Porto Alegre, RS, Brazil}
\author{E.~G.~S.~Luna}
\email{luna@if.ufrgs.br}
\affiliation{Instituto de F\'isica, Universidade Federal do Rio Grande do Sul, Caixa Postal 15051, CEP 91501-970, Porto Alegre, RS, Brazil}

\begin{abstract}
  
We revisit a minijet model to examine the behavior of the total cross section, $\sigma_{tot}$, and the ratio of the real to imaginary parts of the scattering amplitude, $\rho$, at high energies. In this framework, the growth of $\sigma_{tot}$ in $pp$ and $\bar{p}p$  channels is driven by semihard partonic processes dominated by gluon interactions. The QCD contribution to $\sigma_{tot}$ for the jet production is computed in the next-to-leading order. We analyze data separately from the TOTEM and ATLAS/ALFA Collaborations and find evidence in both cases suggesting the necessity of an odd semihard component that becomes asymptotically significant at high energies.
\end{abstract}

\maketitle

\section{Introduction}
In QCD the energy-dependent increase in the $\sigma_{tot}$ during hadronic collisions is attributed to the production of jets with transverse energy $E_{T}$ that is significantly smaller than the total energy $s$ of the collision. These so-called minijets arise from semihard parton-parton interactions, wherein the involved partons carry only small fractions of the momenta of their parent hadrons \cite{ryskin001,ryskin002}. This behavior can be modeled phenomenologically using an eikonal formulation based on QCD, which ensures compliance with principles of analyticity and unitarity \cite{qcdmodel001,luna001,luna002,luna003,luna004,luna005,luna006,luna007,luna008}. By incorporating parton-level cross sections, modern parton distribution functions, and kinematic cutoffs that restrict the analysis to semihard interactions, we can describe the evolution of $\sigma_{tot}(s)$ and the $\rho(s)$ parameter. 

At ultrahigh energies, minijet production is expected to dominate. However, it is assumed that the eikonal function consists of two components: one describing semihard scatterings and another accounting for soft interactions. We analyze $\sigma_{tot}$ and $\rho$ data obtained at LHC energies using a minijet-based formalism while addressing the discrepancies between the measurements reported by the ATLAS/ALFA and TOTEM Collaborations.

\section{The model}

In our model the eikonal function is additive with respect to the soft and the semihard contributions, allowing us to express it as \cite{qcdmodel001,luna001,luna007,luna008}
\begin{eqnarray}
\chi^{\pm}(s,b) = \chi_{_{soft}}^{\pm}(s,b) + \chi_{SH}^{\pm}(s,b).
\end{eqnarray}

The crossing-odd semihard eikonal, $\chi^{-}_{_{SH}}(s,b)$, decreases rapidly with increasing $s$. Thus it is sufficient to set $\chi_{_{SH}}(s,b)=\chi^{+}_{_{SH}}(s,b)$, leading to $\chi^{-}(s,b) = \chi^{-}_{_{soft}}(s,b)$. The even part of the semihard eikonal contribution is given by
\begin{eqnarray}
\chi^{+}_{SH}(s,b) = \frac{1}{2}\, \tilde{\sigma}(s)\, W_{\!\!_{SH}}(b;\nu_{_{SH}}),
\end{eqnarray}
where $\tilde{\sigma}(s)$ is the usual QCD cross section for jet production, and $W_{\!\!_{SH}}(b;\nu_{_{SH}})$ is an overlap density for the partons at $b$,
\begin{eqnarray}
  W_{\!\!_{SH}}(b;\nu_{_{SH}}) &=& \frac{1}{2\pi}\int_{0}^{\infty}dk_{\perp}\, k_{\perp}\, J_{0}(k_{\perp}b) \nonumber \\
  &\times& G_{A}(k_{\perp};\nu_{_{SH}})\,G_{B}(k_{\perp};\nu_{_{SH}}), 
\label{overlap001}
\end{eqnarray}
where $\nu_{_{SH}}$ is a free adjustable parameter, while $G_{A}(k_{\perp};\nu_{_{SH}})$ and $G_{B}(k_{\perp};\nu_{_{SH}})$ are form factors of the colliding hadrons $A$ and $B$. We assume that the parton distribution behaves like a dipole form, specifically
\begin{eqnarray}
G_{A}(k_{\perp};\nu_{_{SH}}) = G_{B}(k_{\perp};\nu_{_{SH}}) = \left( \frac{\nu_{_{SH}}^{2}}{k_{\perp}^{2}+\nu_{_{SH}}^{2}} \right)^{2}.
\label{form09}
\end{eqnarray}
Using this dipole form factor, we can express the overlap density as
\begin{eqnarray}
W_{\!\!_{SH}}(s,b) = \frac{\nu_{_{SH}}^{2}}{96\pi} (\nu_{_{SH}} b)^{3} K_{3}(\nu_{_{SH}} b).
\label{eq20}
\end{eqnarray}

The QCD contribution $\tilde{\sigma}$ to $\sigma_{tot}$ for the process $A+B \to jets$ is given by
\begin{widetext}
\begin{eqnarray}
\tilde{\sigma}(s) &=&  \int_{p_{T_{min}}^{2}}^{s/4} \!\!dp_{T}^{2} \int_{4p_{T}^{2}/s}^{1} \!\!dx_{1}
\int_{4p_{T}^{2}/x_{1}s}^{1} \!\!\!dx_{2} \left[ f_{i/A}(x_{1}, Q^{2})f_{j/B}(x_{2}, Q^{2}) + f_{j/A}(x_{1}, Q^{2})f_{i/B}(x_{2}, Q^{2})
  \right] \nonumber \\
&\times& \left[ \frac{d\hat{\sigma}_{ij\to kl}}{dp_{T}^{2}}(\hat{t}, \hat{u}) +
  \frac{d\hat{\sigma}_{ij\to kl}}{dp_{T}^{2}}(\hat{u}, \hat{t}) \right] \left[ 1\!-\! \frac{\delta_{ij}}{2} \right]\!\! \left[ 1\!-\! \frac{\delta_{kl}}{2} \right] ,
\label{eq08}
\end{eqnarray}
\end{widetext}
where $\hat{s}+\hat{t}+\hat{u}=0$, $\hat{s}=x_{1}x_{2}s$, $\hat{t} = -\frac{\hat{s}}{2} \left( 1 - \sqrt{1 - \frac{4p_{T}^{2}}{\hat{s}}}  \right)$, and $Q^{2} = p_{T}^{2}$. In (\ref{eq08}) $p_{T_{min}}$ denotes the minimal momentum transfer in the semihard scattering, while $x_{1}$ and $x_{2}$ represent the momentum fractions of the parent hadrons $A$ and $B$ carried by the partons $i$, $j$, $k$, and $l$, where $i,j,k,l = q, \bar{q}, g$. The term $\frac{d\hat{\sigma}_{ij\to kl}}{dp_{T}^{2}}$ corresponds to the differential cross section for the scattering process $ij\to kl$, and $f_{i/A}(x_{1},|\hat{t}|)$ $\left[ f_{j/B}(x_{2},|\hat{t})| \right]$ refers to the parton $i$ $\left[ j \right]$ distribution within hadron $A$ $\left[ B \right]$.

As $x\to 0$, the gluon distribution becomes dominant. Consequently, the calculation of $\tilde{\sigma}(s)$ focuses on processes that involve at least one gluon in the initial state, such as $gg \to gg$, $qg\to qg$, $\bar{q}g\to \bar{q}g$, and  $gg\to \bar{q}q$. Given that we employ next-to-leading order (NLO) parton distribution functions (PDFs), the processes $\frac{d\hat{\sigma}_{ij\to kl}}{dp_{T}^{2}}$ must also be evaluated at NLO. To achieve this, we introduce a $K$-factor, defined as the ratio between the NLO and leading-order (LO) cross sections for a specific process. This $K$-factor is incorporated into a phenomenological parameter ${\cal N}$, which additionally accounts for the uncertainty in the choice of $p_{T_{min}}$. As a result, the even part of the semihard eikonal can be effectively represented as
\begin{eqnarray}
\chi^{+}_{SH}(s,b) = \frac{{\cal N}}{2}\, \tilde{\sigma}(s)\, W_{\!\!_{SH}}(s,b).
\end{eqnarray}
The soft eikonal,  required to describe the lower-energy forward scattering data, is given by
\begin{eqnarray}
\chi^{+}_{_{soft}}(s,b) = \frac{1}{2}\, \left[ A + iB + \frac{C}{(s/s_{0})^{\gamma}}\, e^{i\pi\gamma/2} \right] W^{+}_{\!\!_{soft}}(b;\mu^{+}_{_{soft}}), \nonumber \\
\label{soft01}
\end{eqnarray}
\begin{eqnarray}
\chi^{-}_{_{soft}}(s,b) = \frac{1}{2} \frac{D}{\sqrt{s/s_{0}}}\, e^{-i\pi/4}\, W^{-}_{\!\!_{soft}}(b;\mu^{-}_{_{soft}}),
\label{softminus}
\end{eqnarray}
where $\sqrt{s_{0}}\equiv 5$ GeV, and $A$, $B$, $C$, $D$, and $\mu^{+}_{_{soft}}$ are fitting parameters. The parameters $\gamma$ and $\mu^{-}_{_{soft}}$ were assigned the values $0.7$ and $0.5$ GeV, respectively.
The soft form factors are also derived from a dipole basis in $k_{\perp}$. Thus, we have
\begin{eqnarray}
W_{_{soft}}^{i}(b;\mu^{i}_{_{soft}}) = \frac{(\mu^{i}_{_{soft}})^{2}}{96\pi} (\mu^{i}_{_{soft}} b)^{3} K_{3}(\mu^{i}_{_{soft}} b) ,
\label{chDGM.19}
\end{eqnarray}
where $i=+,-$. The forward quantities $\sigma_{tot}(s)$ and $\rho(s)$, expressed in terms of the eikonal function $\chi (s,b)$, are given by
\begin{eqnarray}
  \sigma_{tot}(s) = 4\pi   \int_{_{0}}^{^{\infty}}   \!\!  b\,   db\,
[1-e^{-\chi_{_{R}}(s,b)}\cos \chi_{_{I}}(s,b)] ,
\end{eqnarray}
\begin{eqnarray}
  \rho(s) = \frac{-\int_{_{0}}^{^{\infty}}   \!\!  b\,  
db\, e^{-\chi_{_{R}}(s,b)}\sin \chi_{_{I}}(s,b)}{\int_{_{0}}^{^{\infty}}   \!\!  b\,  
db\,[1-e^{-\chi_{_{R}}(s,b)}\cos \chi_{_{I}}(s,b)]}.
\end{eqnarray}

\section{Results and Conclusions}
To systematically examine the discrepancies between the TOTEM and ATLAS/ALFA results, we conduct global fits to $pp$ and $\bar{p}p$ forward scattering data for $\sqrt{s} \geq 10$ GeV \cite{pdg} using two distinct datasets: one incorporating the ATLAS/ALFA measurements at high energies (defined as Ensemble A) \cite{atlas01,atlas02,atlas03} and the other based on TOTEM measurements at LHC (defined as Ensemble T) \cite{totem01,totem02,totem03,totem04,totem05,totem06}.

With the datasets explicitly defined, we proceed to the phenomenological analysis by performing global fits for the two distinct ensembles. These fits are carried out using a $\chi^{2}$ minimization approach with 90\% CL, where the resulting $\chi^{2}_{min}$ follows a $\chi^{2}$ distribution with $\nu$ degrees of freedom.

For our calculations within the QCD-based framework, we utilized the CT18 parton distribution functions (PDFs) provided by the CTEQ-TEA  collaboration \cite{ct18}. These PDFs are derived from a comprehensive global analysis that incorporates a broad range of high-precision data from the Large Hadron Collider (LHC), the combined HERA I+II deep-inelastic scattering dataset, and the datasets included in the earlier CT14 global QCD analysis.

We achieve a good statistical fit to the overall dataset, as evidenced by the $\chi^{2}/\nu$ value in Table 1. However, the $\rho$ data at $\sqrt{s} = 13$ TeV are not well-described, consistent with expectations based on dispersion relations. These relations connect the central value of $\rho (s)$ to the growth behavior of $\sigma_{tot} (s)$. Consequently, regardless of which normalization of the data proposed by TOTEM or ATLAS/ALFA is accurate, a satisfactory description of the $\rho$ parameter necessitates the inclusion of an odd component in the semihard amplitude that persists at high energies.

As an extension of this study, we aim to examine the impact of an odd semihard component and evaluate the effects of utilizing different sets of parton distribution functions.

\section*{Acknowledgment}

This research was partially supported by the Conselho Nacional de Desenvolvimento Cient\'{\i}fico e Tecnol\'ogico under Grant No. 307189/2021-0.

\begin{figure*}
\label{f1}
\begin{center}
 \includegraphics[scale=0.30]{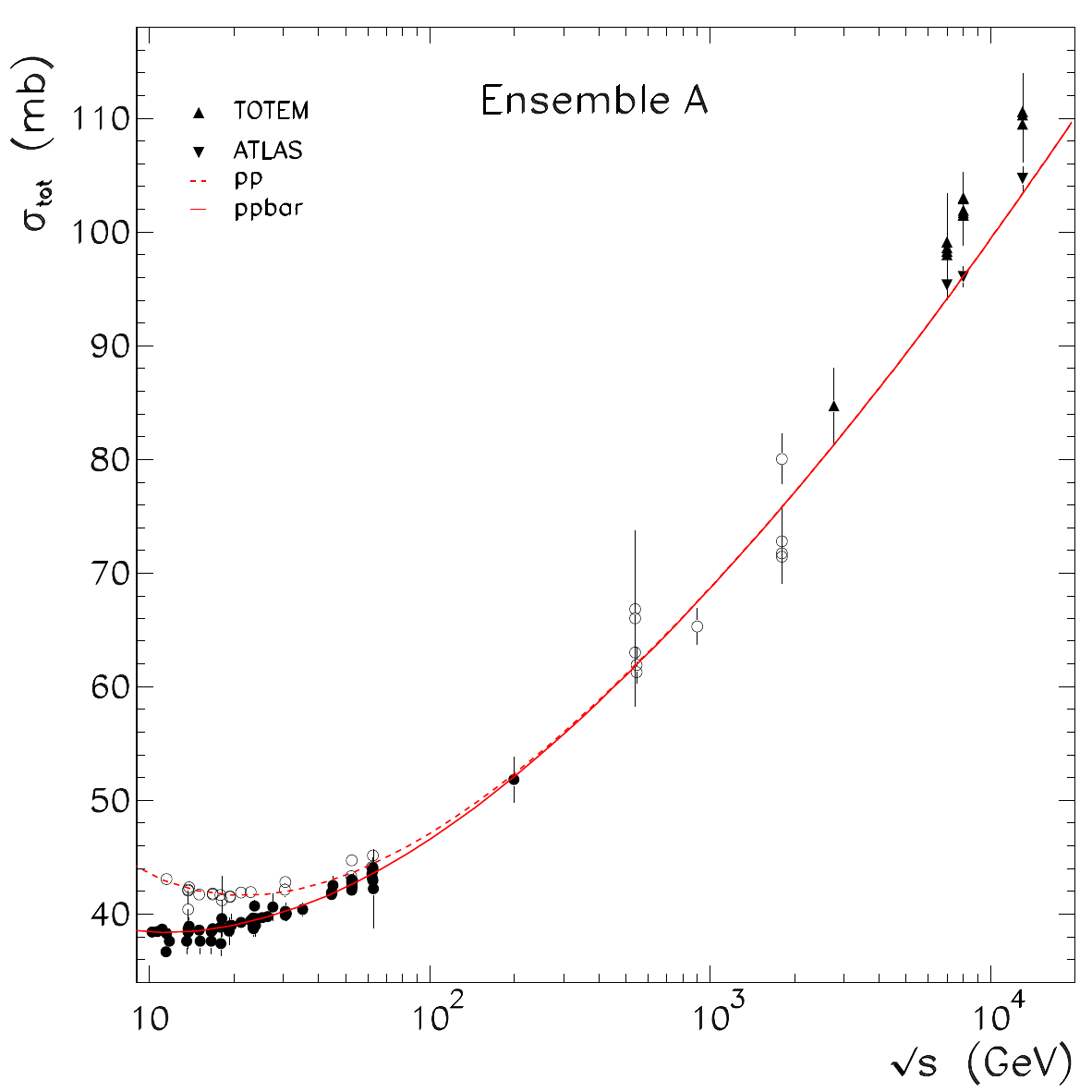}
 \includegraphics[scale=0.30]{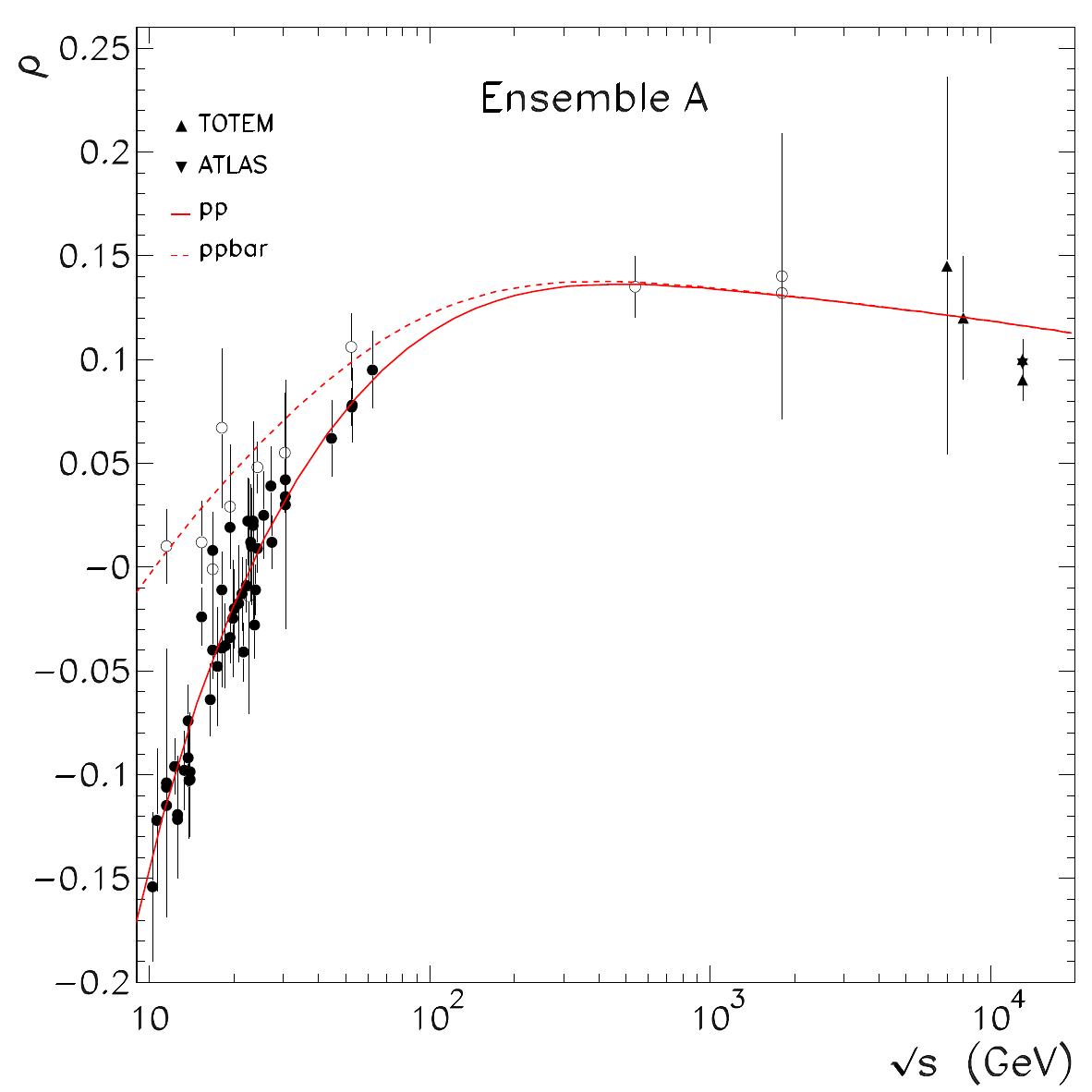}
 \caption{\small The energy behaviour of $\sigma_{tot}$ (left) and $\rho$ (right) from Ensemble A.}
\end{center}
\end{figure*}

\begin{figure*}
\label{f1}
\begin{center}
 \includegraphics[scale=0.30]{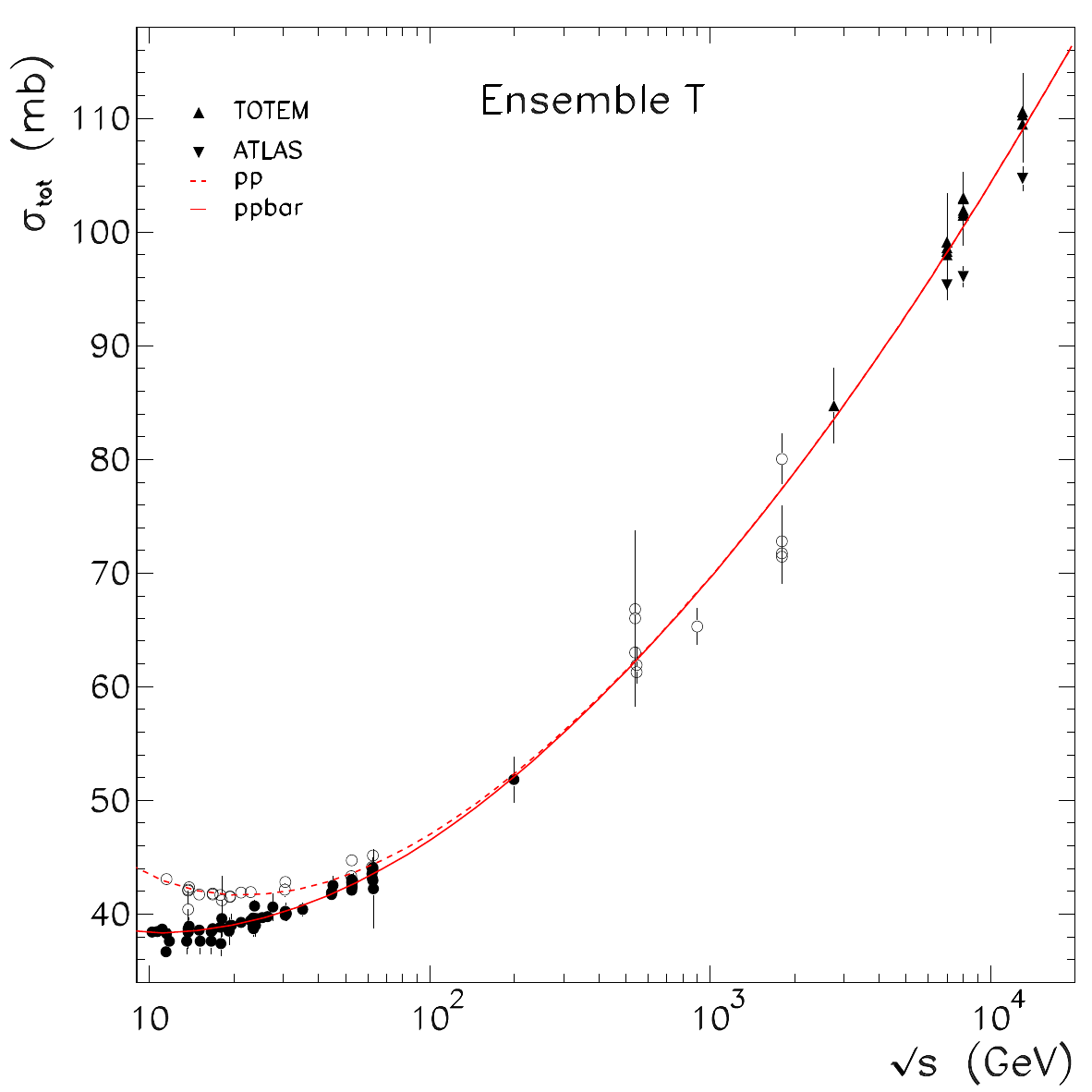}
 \includegraphics[scale=0.30]{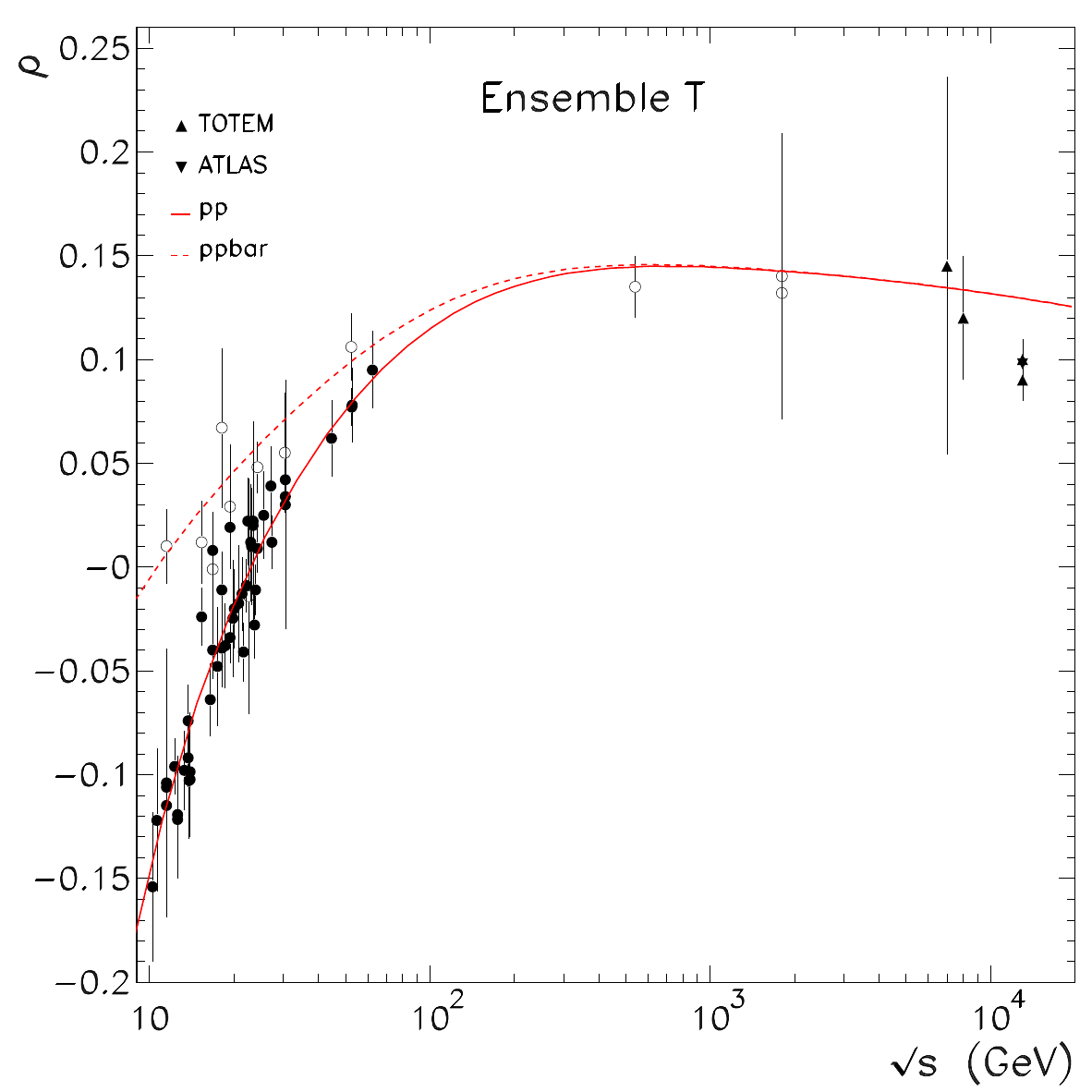}
 \caption{\small The energy behaviour of $\sigma_{tot}$ (left) and $\rho$ (right) from Ensemble T.}
\end{center}
\end{figure*}

\begin{table*}
\centering
\caption{Values of the parameters obtained in the global fits to the Ensemble A and Ensemble T. The data fitting was carried out using $p_{T_{min}} = 1.1$ GeV.}
\vspace{0.3cm}
\begin{tabular}{c|c|c}
\hline\hline 
  & TOTEM & ATLAS \\
\hline 
$\mathcal{N}$  & $1.49\pm 0.50$ & $1.83\pm 0.46$ \\
$\nu_{SH}$ [GeV] & $1.32\pm 0.12$ & $1.42\pm 0.08$ \\
$A$ \, [GeV$^{-2}$]    & $(2.38\pm 0.24)\times 10^{3}$ & $(2.11\pm 2.00)\times 10^{3}$ \\
$B$\, [GeV$^{-2}$]     & $-79.2\pm 130.8$ & $-76.31\pm 148.3$ \\
$C$ \,[GeV$^{-2}$]    & $(33.6\pm 11.2)\times 10^{3}$ & $(31.2\pm 35.5)\times 10^{3}$ \\
$\mu^{+}_{soft}$ [GeV]  & $2.58\pm 0.05 $ & $2.55\pm 0.39$ \\
$D$ \,[GeV$^{-2}$]    & $149.7\pm 11.4$ & $149.5\pm 11.4$ \\
\hline
$\nu$  & $168$ & $158$ \\
$\chi^{2}/\nu$ & $1.18$ & $1.10$ \\
\hline\hline
\end{tabular}
\label{tab001}
\end{table*}

\end{document}